\newcount\pageformat\pageformat=2  
\newcount\pssize

\ifnum\pageformat=2
 \documentclass[aps,prb,twocolumn,floatfix,showpacs,amsmath,amssymb]{revtex4}
\else
 \documentclass[aps,prb,preprint,floatfix,showpacs,amsmath,amssymb]{revtex4}
\fi

\usepackage{psfig}

\def\half{\frac{1}{2}}

\begin{document}


\title{Vortex-vacancy interactions in two-dimensional easy-plane magnets}

\author{G. M. Wysin}
\email{wysin@phys.ksu.edu}
\homepage{http://www.phys.ksu.edu/~wysin/}
\affiliation{Department of Physics, Kansas State University, Manhattan, Kansas 66506--2601}

\date{July 17, 2003}

\begin{abstract}
For a model with isotropic nearest neighbor exchange combined with easy-plane
exchange or single-ion anisotropies, the static effects of a magnetic vacancy
site on a nearby  magnetic vortex are analyzed on square, hexagonal and 
triangular lattices.  
Numerical energy minimization and linear stability analysis using the
vortex instability mode are employed.
When the vortex is centered on a vacancy, the critical anisotropies where the
stable vortex structure switches from out-of-plane to planar form are 
determined, and the vortex energies and magnetizations are found as functions 
of anisotropy.
Consistent with square lattice calculations by Zaspel et al, the strength of
anisotropy needed to stabilize a vortex in the planar form is reduced 
when the vortex is centered on a vacancy, for all three lattices studied.
The vortex-on-vacancy energy is found to be smaller than the typical energy of
a vortex centered between lattice sites in a system without vacancies.
For a vortex separated from a vacancy, the energy found as a function of 
separation demonstrates an {\em attractive} potential between the two.  
\end{abstract}

\maketitle

\section{Introduction}
\label{sec:introduction}
Magnetic vortices in two-dimensional (2D) systems with easy-plane coupling
have attracted a number of investigations related to the structure
and stability of the in-plane and out-of-plane vortex types.
\cite{twovorts,Wysin+88,Gouvea+89,Wysin94,Wysin+98,Zaspel96,Wysin98}
The presence of these two vortex types\cite{twovorts} was found in numerical
simulations to exhibit an instability at a critical 
anisotropy,\cite{Wysin+88,Gouvea+89} at which one type can transform to the 
other due to favorable energetics.
The instability has been explained in terms of the energy balance 
and softening of a dynamic mode on a vortex\cite{Wysin94,Wysin+98}  using
discrete lattice models, with precisely located critical
points found by analyzing progressively larger and larger vortex core regions.
\cite{Zaspel96,Wysin98}
%

A magnetic material is likely to have some distribution of magnetically 
inert or nonmagnetic impurities at some of the lattice sites.  
Effectively these are magnetic vacancies with respect to the magnetic
degrees of freedom.
Each vacancy site is expected to play a role in the structure and 
dynamics of all the magnetic excitations present.
Zaspel et al\cite{Zaspel+96} considered a model for such a nonmagnetic 
impurity in a square lattice 2D ferromagnet (FM).
Applying the same discrete core methods as was done for uniform systems, 
they studied the structure and stability of a vortex centered on a
vacancy site on a square lattice.
They found that the easy-plane anisotropy strength necessary to keep the 
vortex in the in-plane form was drastically reduced by the vacancy.
The effect is significant: the critical anisotropy strength $\delta_{c}$  
(as a fraction of the exchange energy, see definition and Hamiltonian below)  
was found to change from about $0.2966$ for a free vortex to a new value 
$0.0429$ when pinned on a vacancy.
It means that in real materials even the presence of a low density of
vacancies could dramatically influence the static and dynamic
magnetic properties.
Further studies have indicated significant dynamic features attributed 
to vortices pinned on impurities in 
isotropic antiferromagnets (AFM).\cite{Subbaraman+98,Zaspel+02,PerPir03}

On the other hand, M\'ol et al\cite{Mol+02} made a {\em continuum}
calculation for a vortex near a nonmagnetic impurity under the assumption that 
the impurity produces a certain global deformation of the vortex spin structure.
The calculation indicated an effective repulsive potential between a 
planar vortex and the nonmagnetic impurity.
This would suggest that a spin vacancy could not attract or
pin a nearby vortex. 
Then the discrete core model for the calculation of critical anisotropy 
would be of questionable value.
Monte Carlo simulations of a planar rotator model\cite{Leonel+03}
indicated a lowering of the Berezinskii-Kosterlitz-Thouless transition 
temperature\cite{BKT} due to the presence of magnetic vacancies.  
Using a continuum model, Leonel et al\cite{Leonel+03} concluded that
a nonmagnetic impurity will repel either an individual vortex or
even a vortex-antivortex pair, and that this effect can even force the
BKT transition temperature to zero at a critical density of vacancies.
On the other hand, study of a 2D isotropic Heisenberg antiferromagnet
by M\'ol et al\cite{Mol+03} determined oscillatory dynamic modes of
solitons pinned to vacancies, which requires an attractive restoring
potential.
These are contradictory results, therefore, further investigation of
the vacancy--vortex interaction potential is needed.

Although these are related but distinct models (three-component easy-plane,
one-component planar, and AFM vs. FM), it would be satisfying to understand
better the vortex--vacancy interaction problem in general, partly to resolve 
this discrepancy, and to gain a better understanding of
the overall energetics and the influence of different lattices.
The studies here will involve only lattice models; they may give some 
insight when compared to the above mentioned continuum predictions.
Ultimately a lattice model for a vacancy is the original source of any
continuum theoretical description;  the continuum theory cannot completely
describe subtle differences ocurring on the different lattices at short
length scales.
In this regard, the vortex--vacancy interaction is studied here on
2D square, hexagonal and triangular lattices.

For simplicity, we consider the effect of a single vacancy on
a single magnetic vortex.
The static structure and properties of the vortex are found when 
the vortex is centered on the vacancy.
Additionally, an effective vortex-vacancy interaction potential is 
estimated as a function of their separation. 
It is found that the vortex--vacancy interaction potential is always 
attractive, and that at well-chosen anisotropy, the dynamic process 
of pinning of a vortex onto a vacancy could lead to strong out-of-plane 
spin fluctuations at very low energy cost.


\section{Easy-plane model}
The model to be investigated has classical spins defined at lattice
sites $\bf n$ interacting with nearest neighbors at displacements 
$\bf a$ according to the Hamiltonian,
\begin{equation} \label{Ham}
H = J \sum_{\bf n} \Bigl\{ -\half \sum_{\bf a} \Bigl[ 
S^x_{\bf n}S^x_{\bf n+a}+S^y_{\bf n}S^y_{\bf n+a}
+\lambda S^z_{\bf n} S^z_{\bf n+a} \Bigr] 
+ d (S^{z}_{\bf n})^2 \Bigr\}.
\end{equation}
Parameter $\lambda$ determines the exchange anisotropy strength 
$\delta \equiv 1-\lambda$ and $d$ is the single ion anisotropy strength.  
Positive values of $\delta$ and $d$ correspond to the easy-plane anisotropy
that is of interest here.
The $z$-axis is the hard axis; $xy$ is the easy plane.
A ferromagnetic interaction is assumed. 
For static vortex properties, there are essentially no differences 
in energy or critical anisotropies for the antiferromagnetic model on 
square and hexagonal (bipartite) lattices.  
The vortex structure in terms of staggered magnetization for AFM vortices is
the same as that for FM vortices with the same anisotropy parameters.
Obviously these comments do not apply to the AFM model on the triangular lattice,
due to each site having six nearest neighbors, requiring three sublattices,
which results in frustration.
With this exception, study of the FM model is sufficient for determining the
static vortex properties.
Dynamic properties, however, {\em will} be different for AFM vs.\ FM models.

\subsection{Vortices in a Uniform System}
It is convenient to write the spins of length $S$ in terms of in-plane angles 
$\phi_{\bf n}\equiv \tan^{-1}(S^y_{\bf n}/S^x_{\bf n})$
and scaled out-of-plane component $m_{\bf n}\equiv S^z_{\bf n}/S$,
\begin{equation}
\vec{S}_{\bf n}=S(\sqrt{1-m^2_{\bf n}}\cos\phi_{\bf n},
\sqrt{1-m^2_{\bf n}}\sin\phi_{\bf n}, m_{\bf n} ).
\end{equation}
In this notation a static vortex at position $\vec{r}_v=(x_v,y_v)$ in
{\em continuum} theory has in-plane angle in the configuration
\begin{equation}
\label{Phi_in}
\phi(\vec{r}) = 
\Phi^{v}(\vec{r}) = \tan^{-1}\left(\frac{y-y_v}{x-x_v}\right)+\phi_0,
\end{equation}
where $\phi_0$ is an arbitrary constant, set to zero here without
loss of generality.
On a lattice, the in-plane vortex angles $\Phi_{\bf n}^{v}$ lose the perfect 
circular symmetry of this formula, and obtain modifications largest near 
the vortex core.
They satisfy a discrete lattice nonlinear Laplace equation,
\begin{equation}
\label{Laplace}
\sum_{\bf a} \sin(\Phi_{\bf n}^{v}-\Phi_{\bf n+a}^{v}) = 0.
\end{equation}
(The same equation will give the vortex structure also when a vacancy
exists.)
The out-of-plane component $m$ satisfies an appropriate energy minimization
equation resulting from $H$.

It is well known that two vortex types (out-of-plane, at weak anisotropy,
and in-plane, at stronger anisotropy) can be present in this model.
The out-of-plane vortex has a nonzero profile for $m_{\bf n}$ whose
magnitude typically peaks somewhere near the vortex core, and diminishes
over a length scale determined by the inverse anisotropy strength.
The in-plane or planar vortex has $m_{\bf n}=0$ everywhere.
The in-plane vortices require a minimum anisotropy strength (denoted by 
critical parameters $\lambda_c$ or $d_c$ for the different forms of 
anisotropy) in order to be stabilized.
Conversely, the out-of-plane vortices are stable only if the anisotropy
is weaker than the critical value.
For a vortex centered in a plaquette (between lattice sites), the critical 
anisotropy strength for in-plane vortex stabilization increases with the 
coordination number ($z$) of the lattice.\cite{Wysin98}
For hexagonal ($z=3$), square ($z=4$) and triangular ($z=6$) lattices,
the respective critical values are $\delta_c \approx 0.1670, 0.2966, 0.3871$
and $d_c \approx 0.2212, 0.4690, 0.8391$.
Additionally, there is a weak attractive potential pulling the vortex center
to a position of high symmetry at the center of a plaquette.
In the present study, we investigate how the presence of a vacancy 
could modify the critical anisotropies for the different lattices, 
and how a vacancy otherwise affects the vortex structure and energetics.

\section{Vortex on a Vacancy}
\label{sec:critical}
A system is considered with a single magnetic vacancy at its center,
which is taken as the origin of coordinates, $(0,0)$.
The center of the vortex is coincident with the vacant site.
For the purpose of analysis and numerical calculations, a finite
circular system of radius $R$ is considered, on either 
hexagonal (i.e., honeycomb), square or triangular lattices.
This means lattice sites are set up surrounding the origin,
according to the chosen lattice, and only those within radius
$R$ are kept.
The system is abruptly cut off at radius $r=R$; effectively it is
a system with a free boundary condition at this radius.
The missing magnetic site corresponds to a number of missing
exchange bonds equal to the coordination number of the lattice,
$z=3, 4, 6$ for hexagonal, square or triangular lattices.
Thus, the order of magnitude of the energy change due
to a vacancy should be proportional to $zJS^2$.

\subsection{Critical Anisotropy Calculations}
The critical anisotropies are determined by locating the values
where it becomes energetically favorable for the in-plane vortex to
develop nonzero out-of-plane components.
The method used in Refs. \onlinecite{Wysin98}, considering a set of core 
spins that is most important in the energetics, is applied here.
Each spin is assumed to have small amplitude dynamic deviations from the
in-plane vortex structure:
\begin{equation}
\phi_{\bf n}=\Phi_{\bf n}^{v}+\varphi_{\bf n}, \quad S_{\bf n}^{z} = S m_{\bf n}.
\end{equation}
The linearized equations of motion for the fluctuations $\varphi$
and $m$ produce a zero-frequency normal mode when the anisotropy 
strength is at a critical value.
The mode obtains an imaginary frequency when the anisotropy
strength is weaker than the critical value, signifying the 
growth of nonzero out-of-plane spin components, and the
transition to an out-of-plane vortex structure.

\subsubsection{Ring Coordinates on Different Lattices}
This special vortex instability mode of oscillation has a nearly circular 
symmetry around the vortex core, which allows for great simplification 
of its analysis.
In fact, circular symmetry is not exactly valid on a lattice.
Instead, one supposes that a given site with coordinates 
$(x_{\bf n},y_{\bf n})$, and other sites related by symmetry operations
appropriate to that lattice, all have the same in-plane and out-of-plane
spin fluctuations.
These are refered to as ``rings'' in Ref. \onlinecite{Wysin98}. 
All the sites of a given ring have the same radius, however, all sites
at the same radius do not necessarily belong to the same ring.
Also, because the vortex is assumed to be centered on the vacancy,
rather than centered in a plaquette, the ring definitions are different
for the vacancy problem than for the uniform system.

\begin{figure}
\psfig{figure=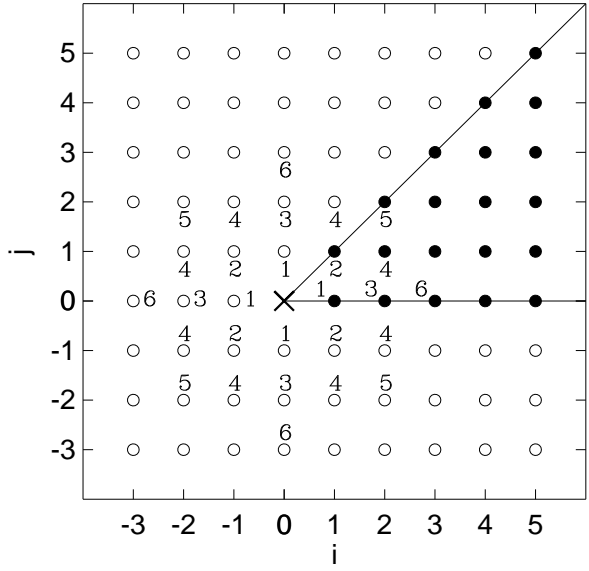,angle=0.0,width=\pssize pt}
\caption{
\label{sqrrings}
Vacancy ($\times$) on a square lattice, and surrounding sites which define the
$(i,j)$ coordinates of the rings (solid points in first octant are the
principal sites).  Numbers indicate the symmetry related sites of some 
of the first few rings.  Rings with principal site on one of the symmetry
lines at $0^{\circ}$ or $45^{\circ}$ have four sites; all others have eight.
}
\end{figure}

On a {\bf square lattice}, the rings can be defined using principal sites 
in the first octant of the $xy$ plane (Fig. \ref{sqrrings}).  
Around the vacant origin,  using lattice constant $a$, a site at 
$(x,y)=a(i,j)$ and all its symmetry related sites at $a(\pm i, \pm j)$ and 
$a(\pm j, \pm i)$ compose a ring.  
A ring $\alpha$ is defined by a pair of positive integers, $\alpha=(i,j)$.
The allowed rings have $i=1,2,3...$ and $j=0,1,2...$ with $j \le i$; 
the restriction $i\ne 0$ produces the vacancy.
The number of members in a ring will be denoted as $\mu_{\alpha}$.
The rings with $j=0$ or $j=i$ lie along lines of high symmetry and
have only $\mu_{\alpha}=4$.
All other rings have $\mu_{\alpha}=8$.
Unlike the uniform system problem, there are no exchange interactions
within an individual ring (no intra-ring exchange energy, $E_{{\rm ee},\alpha}=0$).
For neighboring rings $\alpha=(i,j)$ and $\alpha^{\prime}=(i^{\prime},j^{\prime})$, 
the number of exchange bonds between them is denoted by $c_{\alpha,\alpha^{\prime}}$.
One has $c_{\alpha\alpha^{\prime}}=4=z$ when $j=j^{\prime}=0$ (both
rings' principal sites on the x-axis, a line of high symmetry), and 
$c_{\alpha\alpha^{\prime}}=8=2z$ otherwise.
Before relaxation to a minimum energy in-plane vortex state, the in-plane
vortex angle for principal ring site with $(i,j)$ is simply
$\Phi^{v}_{\alpha}=\tan^{-1}(j/i)$.

\begin{figure}
\psfig{figure=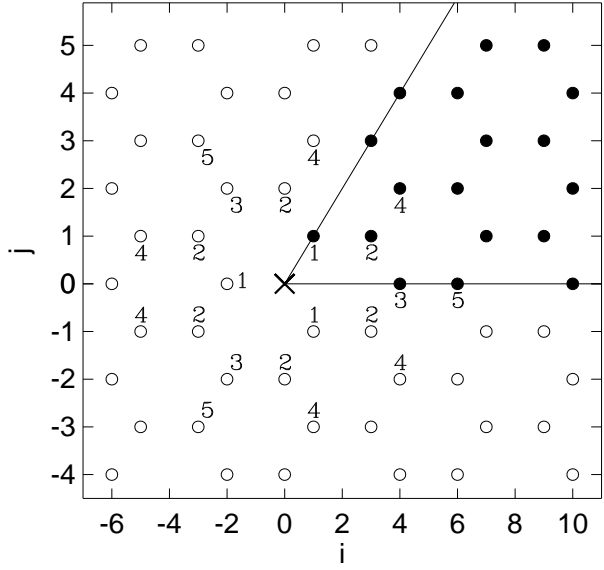,angle=0.0,width=\pssize pt}
\caption{
\label{hexrings}
Vacancy ($\times$) on a hexagonal lattice and definitions of the $(i,j)$ 
coordinates of some of the first few rings.  Rings with principal site on one 
of the symmetry lines at $0^{\circ}$ or $60^{\circ}$ have three sites; all others 
have six.
}
\end{figure}

On a {\bf hexagonal lattice}, a set of principal sites within the first 
sextant of the $xy$ plane define the rings (Fig. \ref{hexrings}).
With the vacancy at the origin, and using integers $i=1,2,3...$ and
$j=0,1,2...$ with $j\le i$, the sites on a ring lie at principal sites
$(x,y)=a(\frac{1}{2}i,\frac{\sqrt{3}}{2}j)$, and other sites related
by $120^{\circ}$ rotations and reflections across the line $i=j$.
The restriction that $(i+j)$ is even produces a triangular lattice 
and the additional restriction $i-3j-2\ne 6n$, $n=$integer, removes 
the hexagon center sites.
One can note that the principal ring sites might also be constructed as the
positions $\vec{r}= l\hat{a}+j\hat{b}$, where $l=\frac{1}{2}(i-j)$,
using basis vectors for hex/tri lattices, $\hat{a}\equiv a(1,0)$ and 
$\hat{b}\equiv a(\frac{1}{2},\frac{\sqrt{3}}{2})$.
The rings along major lines of symmetry where $j=0$ or $j=i$ have
number of members $\mu_{\alpha}=3$;  all other rings have $\mu_{\alpha}=6$.
There is no intra-ring exchange energy, $E_{{\rm ee},\alpha}=0$.
Neighboring rings $\alpha=(i,j)$ and $\alpha^{\prime}=(i^{\prime},j^{\prime})$
along major lines of symmetry $j=j^{\prime}=0$ (both along $\hat{a}$ axis)
or with $(i-j)=(i^{\prime}-j^{\prime})$ (both along $\hat{b}$ axis)
have number of exchange bonds $c_{\alpha\alpha^{\prime}}=3=z$.
Other neighboring rings have $c_{\alpha\alpha^{\prime}}=6=2z$.
Before relaxation, the in-plane vortex angle for principal ring site 
$(i,j)$ is $\Phi^{v}_{\alpha}=\tan^{-1}(\sqrt{3}j/i)$.

\begin{figure}
\psfig{figure=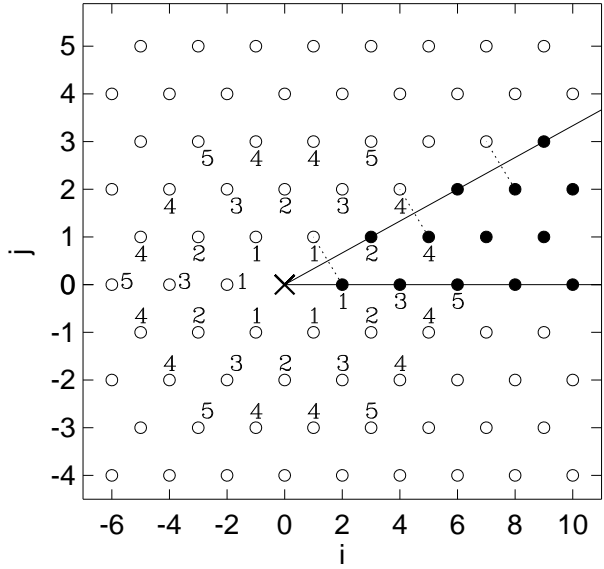,angle=0.0,width=\pssize pt}
\caption{
\label{trirings}
Vacancy ($\times$) on a triangular lattice and definitions of the $(i,j)$ 
coordinates of some of the first few rings.  Rings with principal site on one 
of the symmetry lines at $0^{\circ}$ or $30^{\circ}$ have six sites; all others
have twelve. The dashed lines exhibit some of the intra-ring or self-exchange 
bonds.
}
\end{figure}

For the {\bf triangular lattice} system, a set of principal sites
in the first $30^{\circ}$ above the $x$-axis is sufficient to define
the rings (Fig. \ref{trirings}).
One needs ring sites as on the hexagonal lattice, with $(x,y)=
a(\frac{1}{2}i,\frac{\sqrt{3}}{2}j)$, where $i=1,2,3...$ and
$j=0,1,2...$ with the restriction that $(i+j)$ is even.
To constrain to the principal sites in the first $30^{\circ}$ slice 
requires $j\le \frac{1}{3}i$.  
The rings with principal site along high symmetry lines at 
$0^{\circ}$ ($j=0$) and $30^{\circ}$ ($j=\frac{1}{3}i$) to the $+x$-axis 
have number of members $\mu_{\alpha}=6$; all others have $\mu_{\alpha}=12$.
The rings such as $(i,j)=(2,0), (5,1), (8,2),$ etc., whose principal 
sites lie just below the $30^{\circ}$ line, satisfying $j=\frac{1}{3}(i-2)$, 
have six intra-ring exchange bonds across the $30^{\circ}$ line.
The intra-ring exchange energy in one of these rings with principal site in-plane 
vortex angle $\Phi^{v}_{\alpha}$ and out-of-plane fluctuation $m_{\alpha}$ is
\begin{equation}
E_{{\rm ee},\alpha} \equiv 
-6JS^2 \left\{ (1-m_{\alpha}^2)\cos[2(30^{\circ}-\Phi_{\alpha})]
+\lambda m_{\alpha}^2 \right\}.
\end{equation}
Just as for the hexagonal lattice, neighboring rings 
$\alpha=(i,j)$ and $\alpha^{\prime}=(i^{\prime},j^{\prime})$
along major lines of symmetry $j=j^{\prime}=0$ 
or with $(i-j)=(i^{\prime}-j^{\prime})$ 
have $c_{\alpha\alpha^{\prime}}=6=z$, and other neighboring
rings have $c_{\alpha\alpha^{\prime}}=12=2z$.
The unrelaxed in-plane vortex angle for principal ring site
$(i,j)$ is $\Phi^{v}_{\alpha}=\tan^{-1}(\sqrt{3}j/i)$.
%

\subsubsection{Relaxed Static In-Plane Vortices}
It was mentioned in the Introduction that the static in-plane angles 
satisfy the discrete nonlinear Laplace equation (\ref{Laplace}),
even with a vacancy in the system.  
This is seen even from minimizing the total energy in the 
ring coordinates, under the stipulation that all out-of-plane
fluctuations $m_{\alpha}=0$; see the Hamiltonian in the next section.
Therefore, before proceeding to get critical anisotropies, the
in-plane vortex structure, as defined using the ring variables, was 
relaxed to a local minimum energy configuration, consistent with 
equation (\ref{Laplace}).
This was achieved by an iteration procedure in which each ring 
angle $\Phi^{v}_{\alpha}$ was adjusted to point along the effective
field of its neighboring rings, scanning through all the active rings,
until the magnetization and energy changes fell below a desired precision.
Typically, such a relaxation procedure only makes very minor changes
(of the order of 1\% or less) in the spin components at sites close 
to the vortex core.
Farther from the core, the relaxed spin angles are well-described by the
continuum formula, Eq.\ (\ref{Phi_in}).

\subsubsection{Zero-Frequency Dynamic Mode}
Based on the definitions of the rings, it is straightforward to 
determine which pairs of rings are neighbors,
denoting them as $(\alpha,\alpha^{\prime})$.
Assuming a symmetrical dynamical mode of oscillation, where the
spin fluctuations are the same for all sites on a ring (FM model),
the total system Hamiltonian is written $H=H_{\rm int}+H_{\rm self}$, 
where the ring {\it interaction} and {\it self} energies are
\begin{eqnarray}
\label{int}
H_{\rm int} &=& -JS^2 \sum_{(\alpha,\alpha^{\prime})}
c_{\alpha\alpha^{\prime}} 
\Bigl[ 
\lambda m_{\alpha}m_{\alpha^{\prime}}  \nonumber \\
&+&  \sqrt{1-m_{\alpha}^2}\sqrt{1-m_{\alpha^{\prime}}^2}
\cos(\Phi^{v}_{\alpha}-\Phi^{v}_{\alpha^{\prime}}) 
\Bigr],
\end{eqnarray}
\begin{equation}
\label{self}
H_{\rm self}= \sum_{\alpha} \left[ E_{{\rm ee},\alpha} 
+ dJS^2 \mu_{\alpha} m_{\alpha}^2 \right].
\end{equation}
The sum in (\ref{int}) is only over the neighboring pairs of rings,
and the sum in (\ref{self}) is over all rings.

The classical dynamics for the in-plane spin fluctuations follows 
from\cite{Wysin98}
\begin{equation}
S\mu_{\alpha}\dot{\varphi}_{\alpha} = \frac{\partial H}{\partial m_{\alpha}}
\end{equation}
with a similar equation for the conjugate coordinate $m_{\alpha}$.
This equation is sufficient to determine when a zero-frequency 
mode occurs, locating a critical anisotropy.  
The equation can be written in a matrix form,
\begin{equation}
\dot{\varphi}_{\alpha} = \sum_{\alpha^{\prime}} 
{\cal F}_{\alpha,\alpha^{\prime}} m_{\alpha^{\prime}}.
\end{equation}
The mode frequency goes to zero when the determinant of $\cal F$
goes to zero, which gives the condition for finding a critical anisotropy. 
This was performed by setting up the matrix $\cal F$ numerically
for a number of rings $N$, and using a secant search method, 
adjusting $\lambda$ or $d$ while holding the other fixed, 
until a zero determinant resulted.  

\subsubsection{Critical Anisotropies--Numerical Results}
\label{critnum}
For a total number of rings $N\le 3$, one can easily write down explicitly
the total ring energy and matrix ${\cal F}$, and determine critical
anisotropies analytically, as was done by Zaspel et al\cite{Zaspel96}
for the square lattice vacancy problem.
The vacancy-influenced critical exchange anisotropy 
$\delta_{cv}=1-\lambda_{cv}$ is found
for null single-ion anisotropy, $d=0$.
The vacancy-influenced critical single-ion anisotropy $d_{cv}$ is found 
for null exchange anisotropy, $\delta=0$ or equivalently, $\lambda=1$.
The results from a C-program for the $N$-ring calculation agreed
with the analytic calculations for $N\le 3$ on all three lattices,
verifying the reliability of the program.
Obviously, the critical values depend on $N$, therefore, results
were obtained for a range of $N$ up to the order of 1000 rings
until convergence of $\lambda_{cv}$ and $d_{cv}$ to 10 significant figures.
A smaller number of rings ($N\approx 400$) is necessary to achieve this 
for the triangular lattice, probably because it is denser.
A summary of the results is shown in Tables \ref{hexdata},
\ref{sqrdata}, and \ref{tridata}.
On the square and hexagonal lattices, the calculation gives no solution
for either critical anisotropy at $N=1$.
On all the three lattices, a minimum number of rings is necessary
before the critical anisotropies even move into the easy-plane
range ($\lambda_{cv}<1$ and $d_{cv}>0$).

The converged critical exchange anisotropies are $\delta_{cv} \approx
0.0261392320, 0.0455022615, 0.0606587833$, for hexagonal, square and
triangular lattices, and the corresponding single-ion critical
anisotropies are $d_{cv} \approx 0.03674492096, 0.0824221891, 0.1608450953$.
Just as found for a vortex centered in a plaquette, the critical
anisotropies increase with the coordination number, and the single-ion
critical value is larger than the exchange value. 
The results confirm and expand upon those found by Zaspel et al.\cite{Zaspel96}
Most importantly, a vortex on a vacancy will tend to be stabilized  
in the planar form, even at fairly weak anisotropy that would
otherwise produce stable out-of-plane vortices far from any vacancies.
The numerical values for the square lattice are slightly different here
due to using the energetically relaxed in-plane vortex profile
(not Eq. \ref{Phi_in}), and, calculations to a large
number of rings until convergence, rather than an extrapolation
procedure from a small number of rings.
%


\begin{table}
\caption{
\label{hexdata}
{\bf Hexagonal lattice} critical anisotropy parameters for a relaxed vortex 
on a vacancy as a function of the number of rings $N$ and system radius $R$.
}
\smallskip
\begin{tabular}{cccc}
$N$ & $R/a$ & $\lambda_{cv}$ & $d_{cv}$  \cr
\hline
   2  &  1.732 & 1.5636211762  & -0.38565740698 \cr
   3  &  2.000 & 1.2077648210  & -0.20076947906 \cr
   4  &  2.646 & 1.1101053643  & -0.11918190977 \cr
  10  &  4.359 & 1.0023063573  & -0.00297428015 \cr
  11  &  4.583 & 0.9962392581  &  0.00492191196 \cr
  20  &  6.245 & 0.9800726402  &  0.02721911999 \cr
  100 & 15.000 & 0.9738836510  &  0.03669965239 \cr
  400 & 30.643 & 0.9738607689  &  0.03674491810 \cr
 1500 & 60.108 & 0.9738607680  &  0.03674492096 \cr
 3000 & 85.497 & 0.9738607680  &  0.03674492096 \cr
\end{tabular}
\end{table}

\begin{table}
\caption{
\label{sqrdata}
{\bf Square lattice} critical anisotropy parameters for a relaxed vortex on a 
vacancy as a function of the number of rings $N$ and system radius $R$.
}
\smallskip
\begin{tabular}{cccc}
$N$ & $R/a$ & $\lambda_{cv}$ & $d_{cv}$  \cr
\hline
   2 &  1.414 & 1.4137593264 & -0.4065927450 \cr
   3 &  2.000 & 1.2807409281 & -0.3077361144 \cr
   4 &  2.236 & 1.0608891919 & -0.0871050739 \cr
   6 &  3.000 & 1.0227216056 & -0.0346427396 \cr
   7 &  3.162 & 0.9944246293 &  0.0089633765 \cr
  10 &  4.123 & 0.9722646499 &  0.0469741982 \cr
 100 &  15.00 & 0.9544993839 &  0.0824157890 \cr
 400 &  30.89 & 0.9544977384 &  0.0824221893 \cr
 800 &  44.05 & 0.9544977385 &  0.0824221891 \cr
1500 &  60.80 & 0.9544977385 &  0.0824221891 \cr
\end{tabular}
\end{table}

\begin{table}
\caption{
\label{tridata}
{\bf Triangular lattice} critical anisotropy parameters for a relaxed vortex 
on a vacancy as a function of the number of rings $N$ and system radius $R$.
}
\smallskip
\begin{tabular}{cccc}
$N$ & $R/a$ & $\lambda_{cv}$ & $d_{cv}$  \cr
\hline
   1  &  1.000 & 1.8660254038 & -0.8660254038 \cr
   2  &  1.732 & 1.2709004131 & -0.4284919249 \cr
   3  &  2.000 & 1.0669263945 & -0.1342143055 \cr
   4  &  2.646 & 0.9969826115 &  0.0068108464 \cr
  10  &  4.583 & 0.9454955504 &  0.1397956043 \cr
 200  &  24.43 & 0.9393412168 &  0.1608450949 \cr
 400  &  35.04 & 0.9393412167 &  0.1608450953 \cr
 800  &  50.21 & 0.9393412167 &  0.1608450953 \cr
\end{tabular}
\end{table}

\subsection{Relaxed Vortices' Energy and Magnetization}
\label{relaxed}
It is interesting to confirm the locations of the critical anisotropies
by analyzing the total energy and total out-of-plane magnetic moment
of a relaxed vortex, both as functions of anisotropy.
A vortex is considered as above, at the center of a circular system
centered on a vacancy. 
The in-plane and out-of-plane spin components both were allowed to vary, 
until a local minimum energy configuration for $H$ was obtained.
In actual practice, the minimizing configuration was found by
using the original lattice spin fields $\vec{S}_{\bf n}$
and iteratively repointing each along the effective local field 
$\vec{F}_{\bf n}$ due to its neighbors, 
\begin{equation}
\label{Field}
\vec{F}_{\bf n} \equiv -\frac{\partial H}{\partial \vec{S}_{\bf n}}
 = J \left[ \sum_{\bf a} 
\left( 
\vec{S}_{\bf n+a}+\delta ~ S^{z}_{\bf n+a}\hat{z}
\right)
+ 2d ~ S^{z}_{\bf n}\hat{z} \right]
\end{equation}
Scanning linearly through the lattice, each site was updated in sequence,
being reset along the net field due partly to some unchanged neighbors 
and some that have already been repointed. 
This gives slightly faster convergence than a synchronized global update.

Due to the planar symmetry of $H$, if all spins are initially set in the 
$xy$ plane, they will remain precisely in the $xy$ plane under this algorithm, 
even in a situation where, for example, a planar vortex would be unstable.
Therefore, in order to allow for the possibility to relax either to a planar 
or out-of-plane vortex form, nonzero initial values $S^{z}_{\bf n}=10^{-3}$ 
were set, together with initial in-plane vortex angles given by 
Eq.\ (\ref{Phi_in}).
In situations where an in-plane vortex is stable, the $S^z$ spin components
decay away to zero; conversely, at anisotropy where an out-of-plane 
vortex is stable, the $S^z$ components grow into the appropriate out-of-plane
vortex profile.
The iterations were allowed to proceed until the average spin changes 
per site fell below a desired level, typically on the order of 1 part in 
$10^{16}$.
By the time the individual spin changes reached this size, the energy
converged to an even higher precision. 
It is interesting also to note that this iteration converges all the
more slowly as the anisotropy comes closer to a critical value, a 
characteristic feature related to the slow dynamics of the soft mode 
responsible for the in-plane to out-of-plane instability.
 
In a finite system, the boundary can either enhance or diminish the ability
of nearby spins to tilt out of the easy plane, affecting the calculated
critical anisotropy values, as seen above.
For all three lattices studied, fairly good convergence of critical values
took place below a system radius $R=50 a$.  
Therefore we calculated relaxed vortex states for systems with $R= 50 a$;
the spins at the edge of the system have no constraint from outside, that is,
a free boundary condition holds.
The energy was calculated relative to the ground state energy for
spins aligned within the $xy$ plane, an amount of $-JS^2$ per exchange bond.
In addition to the energy, the total magnetic moment was calculated,
having only a $z$ component: $M = \sum_{\bf n} S^{z}_{\bf n}$.
Its deviation from zero gives a distinct signature for the transition
to an out-of-plane state.

\begin{figure}
\psfig{figure=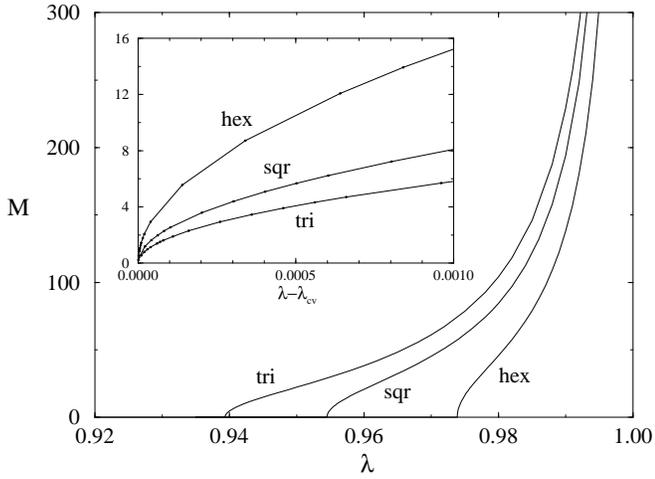,angle=-90.0,width=\pssize pt}
\caption{
\label{MGraf}
Total out-of-plane magnetization of a numerically relaxed vortex-on-vacancy 
versus exchange anisotropy constant $\lambda$, for hexagonal, square and 
triangular lattice systems of radius $R=50a$ with free boundaries.   
The inset shows the same data versus shifted exchange anisotropy 
constants $\lambda-\lambda_{cv}$ near the critical points.
}
\end{figure}

\begin{figure}
\psfig{figure=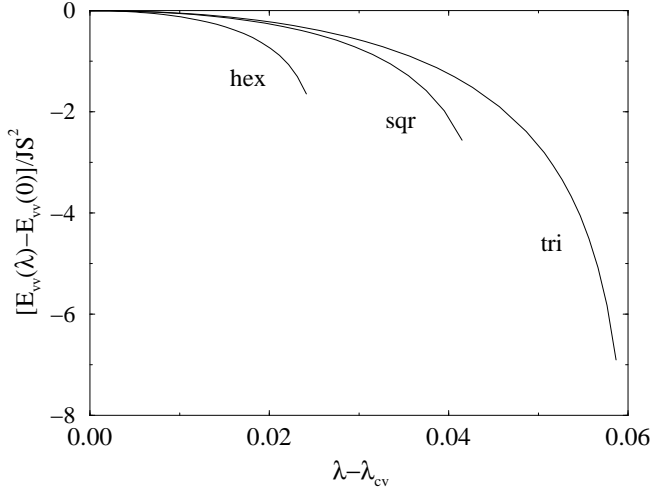,angle=-90.0,width=\pssize pt}
\caption{
\label{EGraf}
Energy of a numerically relaxed vortex-on-vacancy relative to in-plane
vortex-on-vacancy energies $E_{\rm vv}(0)$ versus shifted exchange anisotropy 
constant $\lambda-\lambda_{cv}$, for hexagonal ($E_{\rm vv}(0)=7.143 JS^2$), 
square ($E_{\rm vv}(0)=13.15 JS^2$) and triangular 
($E_{\rm vv}(0)= 23.46 JS^2$) lattice systems of radius $R=50a$ with 
free boundaries.
}
\end{figure}

\begin{figure}
\psfig{figure=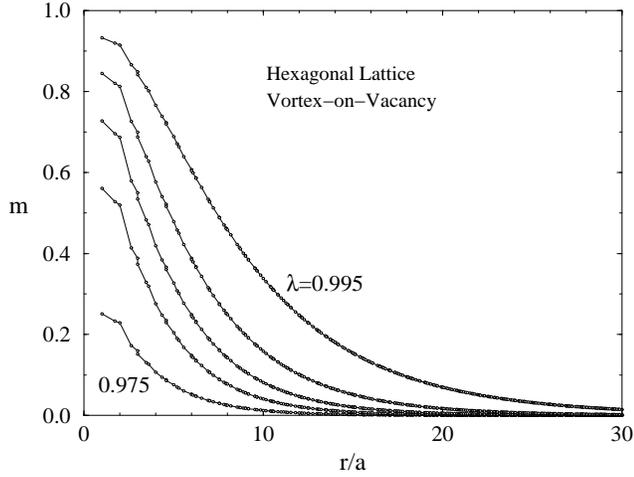,angle=-90.0,width=\pssize pt}
\caption{
\label{Profiles}
Radial dependences of the out-of-plane component of numerically relaxed 
vortex-on-vacancy structures,  at $\lambda=0.975, 0.980, 0.985, 0.990, 0.995$, 
on a hexagonal lattice system of radius $R=50a$ with free boundaries.
}
\end{figure}

Results are presented here using exchange anisotropy only, setting the
single-ion anisotropy to zero.
The vortex magnetic moments $M$ found as a function of $\lambda$ are shown
in Fig.\ \ref{MGraf}.
The critical anisotropies obtained by numerical relaxation of the vortex
structure, and observed in this graph, confirm to several 
digits the critical anisotropies determined in Sec.\ \ref{critnum}.
The same data is replotted in the inset of Fig. \ref{MGraf} versus scaled 
anisotropy $\lambda-\lambda_{cv}$, using the precise critical values from Sec.\
\ref{critnum}.  
One sees that once $\lambda$ passes $\lambda_{cv}$, the magnetic moment grows 
initially proportional to $\sqrt{\lambda-\lambda_{cv}}$, and fastest on
the more open hexagonal lattice, and slowest on the denser triangular lattice.
There is a corresponding extremely soft change in the total system energies 
(relative to the planar vortex energy $E_P$) as
shown in Fig. \ref{EGraf}, verifying that the out-of-plane transition
is towards a vortex of lower energy, once the anisotropy becomes
weaker than the critical strength.  
Some typical out-of-plane vortex profiles are exhibited in Fig. 
\ref{Profiles}, indicating not only the growth of total $M$ with deviation 
from critical anisotropy, but also an associated increase in the vortex
radial length scale as the anisotropy {\em strength} diminishes.
Figure \ref{Profiles} also lends support for Zaspel et al's explanation for 
the lowering of the critical anisotropy strengths due to the presence of 
the vacancy.    
The missing bonds reduce the total energy of both types of vortices, however,
the in-plane vortex energy is reduced more than the typical out-of-plane
vortex energy (except at rather weak anisotropy) because the (removed) 
out-of-plane tilted exchange bonds were relatively smaller energy 
contributions than if they had been lying purely in-plane. 
This is true on all the lattices studied, the differences between them
being caused by the different densities (sites per unit area).
%

\section{Vortex-Vacancy Interaction Potential}
\label{sec:potential}
The above analysis requires that a vortex is actually attracted
energetically to a vacancy.  
This is a reasonable supposition; the missing bonds clearly reduce the
total system energy, and especially the exchange energy due to the in-plane
components if the vortex is centered on the vacancy, because the
spins near a vortex core are the most strongly misaligned ones.
A lesser reduction in in-plane exchange energy results if the vortex is
near the vacancy but not exactly centered on it.
Therefore one can conclude that there should be an attractive 
potential pulling a vortex into a vacancy site.
(M\'ol et al\cite{Mol+02} came to a different conclusion, estimating a
repulsive potential between a vortex and vacancy in a continuum model.)
Due to its intrinsic interest as well as the implications for the
acceptable interpretation of the vacancy modified critical anisotropies,
an analysis of the interaction potential between a vacancy and
a vortex (in a lattice system) is called for.
Therefore we investigate further the vortex-vacancy energetics,
especially as a function of their separation.
%

\subsection{Vortex-on-Vacancy Pinning Energy}
\label{DeltaE}
\begin{figure}
\psfig{figure=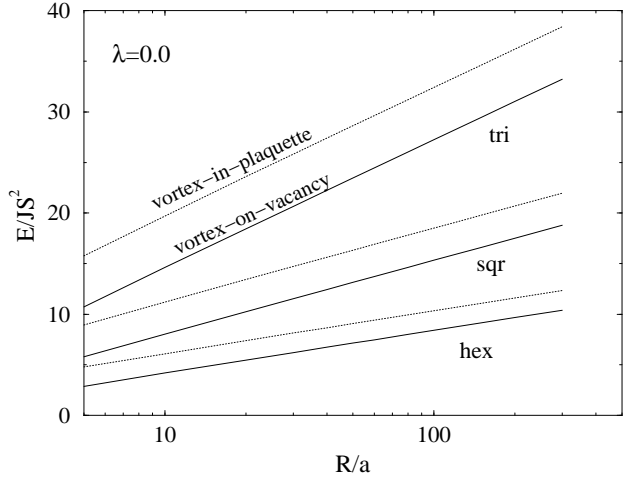,angle=-90.0,width=\pssize pt}
\caption{
\label{EvsR0}
Comparison of vortex-in-plaquette energies (dotted) with vortex-on-vacancy energies
(solid) for in-plane vortices ($\lambda=d=0$) on circular systems of radius $R$,
showing a substantial pinning energy.
}
\end{figure}

As a first step, which suggests clearly that the potential should
be attractive, a comparison is made of the energies of a single
vortex at the center of a circular system of radius $R$, in two 
obvious cases: 1) the lattice is uniform (no vacancy) and the vortex 
is centered in a plaquette; 2) the vortex is centered on a vacancy
at the system center. 
By using the energy minimizing iterative method [Eq.\ (\ref{Field})] just 
described, the vortex configurations for the two cases are easily obtained,
and their energies can be compared.  
At the same radii $R$, the two defined systems do not necessarily have the 
same number of lattice sites or bonds (their lattices are shifted relative
to one another and the circular boundary), so the comparison may require
careful interpretation.
We proceeded by producing a set of energies as a function of system size 
and compare the two results versus $R$.
In Fig. \ref{EvsR0} the comparison is made for {\em in-plane} vortices on
systems with $\lambda=0$ and $d=0$; the vortex-on-vacancy energy
lies below the vortex-in-plaquette energy for all radii. 
The difference is the vortex-on-vacancy binding energy or pinning energy,
$\Delta E$.
For example, for in-plane vortices on the {\bf square lattice}, 
$\Delta E_{\rm sqr} \approx 3.178 JS^2$, more or less independently 
of the system size, and independent of the anisotropy as long 
as the vortices are both of the in-plane type.
The in-plane vortices on {\bf hexagonal} and {\bf triangular} 
lattices indicate binding energies of $\Delta E_{\rm hex} \approx 1.937 JS^2$ 
and $\Delta E_{\rm tri} \approx  5.174 JS^2$, respectively (for $\lambda=d=0$). 
The conclusion is obvious, that if observed in an extended system (aside 
from possibly strong boundary effects), an in-plane vortex {\em will} lower 
its energy by ocurring centered on a vacancy, rather than in a plaquette.

\begin{figure}
\psfig{figure=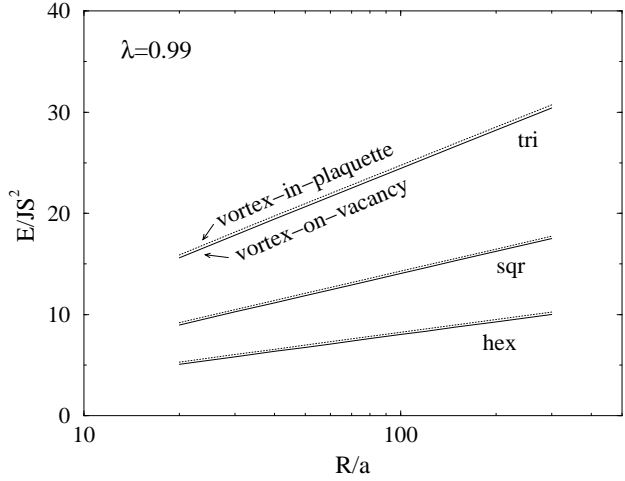,angle=-90.0,width=\pssize pt}
\caption{
\label{EvsR99}
Comparison of vortex-in-plaquette energies (dotted) with vortex-on-vacancy energies
(solid) for out-of-plane vortices ($\lambda=0.99$, $d=0$) on circular systems of 
radius $R$, showing a very weak pinning energy (See Table \protect\ref{Reduction}.).
}
\end{figure}

In Fig. \ref{EvsR99}, a typical comparison is made for {\em out-of-plane 
vortices}, using $\lambda=0.99$ (greater then $\lambda_{cv}$ on all
lattices studied) and $d=0.0$; again the vortex-on-vacancy energy 
is lower for all system sizes.
(The calculation required a larger system than for $\lambda=0$ in order
to produce stable out-of-plane vortex profiles in a finite radius.)
For the square lattice, the vortex binding energy is reduced drastically, 
to about $ 0.232 JS^2$; this value is expected to be a function of the 
anisotropy parameters.
One could also consider an alternative situation where the vortex is
of out-of-plane form when centered in a plaquette, but destabilizes to
in-plane form if centered on a vacancy (whenever 
$\lambda_c < \lambda < \lambda_{cv}$,  such as any of
the three lattices studied with $\lambda=0.93$ and $d=0$).
In this latter example, the vortex-on-vacancy energy is
again consistently lower than the vortex-in-plaquette energy at the
same radius, indicating once again an energetic attraction to 
a vacancy.
In this case the pinning process also will be associated with a significant
signature: the complete {\em elimination} of the out-of-plane magnetization 
component. 

\begin{figure}
\psfig{figure=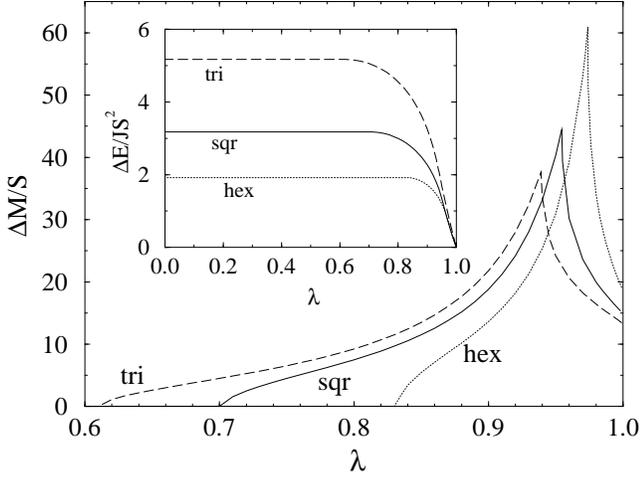,angle=-90.0,width=\pssize pt}
\caption{
\label{Pinning}
The changes in vortex out-of-plane magnetic moment ($\Delta M$) 
and in vortex energy ($\Delta E$, in inset) when a vortex is pinned 
on a vacancy (changing from centered in a plaquette to centered on a
vacancy), as functions of easy-plane anisotropy parameter on
hexagonal, square and triangular lattices. 
}
\end{figure}

Binding energies for vortex-on-vacancy pinning are given in 
Table \ref{Reduction}, for the three anisotropy parameters 
$\lambda=0,$ 0.93, and 0.99, corresponding to permanently in-plane, 
out-of-plane transforming to in-plane, and permanently
out-of-plane vortices.
The  decrease ($\Delta M$) in total out-of-plane magnetic moment as a 
vortex is pinned on a vacancy is also summarized there.
($\Delta M$ is the difference of the vortex-in-plaquette 
and the vortex-on-vacancy magnetic moments.)
Pinning energies and magnetic moment changes were calculated for 
systems with radius $R=100a$, for the full range of easy-plane
exchange anisotropy, as displayed in Fig.\ \ref{Pinning}.
(As explained above, at this radius, the results do not have
any dependence on system size.)
At weak anisotropy ($\lambda \to 1$ ) the pinning energy increases
almost linearly with anisotropy strength.
The pinning energy saturates at its largest value once $\lambda<\lambda_c$
for the lattice under study; the pinning energy is always highest
for the more dense lattices.
The magnetization change on pinning is even more interesting.
It starts at a nonzero value for the isotropic limit ($\lambda=1$),
reaching a sharp maximum exactly at $\lambda_{cv}$ for the lattice
under study (transition between free out-of-plane vortex and pinned
in-plane vortex).
Finally $\Delta M$ becomes zero at $\lambda_c$, where the pinning 
involves only planar vortices.

The region where the pinning leaves the vortices in out-of-plane form 
($\lambda>\lambda_{cv}$) is of special interest;  the pinning is 
associated with relatively small energy change in combination with a 
significant magnetization change.
Such a vortex pinning/depinning transition might be initiated due to thermal 
fluctuations under appropriate conditions, associated with large out-of-plane
magnetization fluctuations.
One the other hand, close to the planar limit ($\lambda=0$), the large
pinning energy means that vortices formed thermally will tend to pin on any 
vacancies present.
Free vortex density could be greatly reduced if there is high enough
vacancy density.

\begin{table}
\caption{
\label{Reduction}
Some vortex-on-vacancy binding energies (in units of $JS^2$) and magnetization 
reductions (in units of $S$) estimated by comparing to the 
vortices-in-plaquette structures ($d=0$), using circular systems of 
radius $R\le 300a$.  
}
\smallskip
\begin{tabular}{ccccccc}
$\lambda$ & $\Delta E_{\rm hex}$ & $\Delta M_{\rm hex}$ & 
            $\Delta E_{\rm sqr}$ & $\Delta M_{\rm sqr}$ & 
            $\Delta E_{\rm tri}$ & $\Delta M_{\rm tri}$ \cr 
\hline
0.0  & 1.937  & 0.0    & 3.178  & 0.0    & 5.174  & 0.0 \cr
0.93 & 1.486  & 21.31  & 1.807  & 28.05  & 2.45   & 32.42  \cr
0.99 & 0.224  & 24.63  & 0.232  & 17.27  & 0.310  & 14.85  \cr
\end{tabular}
\end{table}

\subsection{Separated Vortex--Vacancy Potentials}
To proceed further, it is necessary to study the energy of a vortex 
separated from a vacancy by some distance $r_{\rm vv}$.
The energy minimization scheme of Sec.\ \ref{relaxed} is to be
applied to different vortex-vacancy arrangements.
The potential is considered here for square, hexagonal and triangular
lattices.
Generally, the interaction potential between the two could depend on the 
direction relative to the lattice, as well as the radial separation.
%

In order to investigate this, again a circular system of
radius $R$ is used, with a vortex placed at its center, which is
the origin of coordinates, $(0,0)$.
The vacancy is placed at some position $\vec{X}=(X_x,X_y)$ which is
allowed to vary. 
Alternatively, if the vortex were placed at different positions
away from the system center, a boundary energy that changes significantly 
with the vortex position would result, regardless of the presence or 
absence of a vacancy.
To avoid this complication, it is much simpler to fix the vortex
position at the system center, and move the vacancy around, measuring
the total system energy.
It is not expected that there is a strong energy change due to the
vacancy approaching the system boundary.
Therefore, one expects the resulting total energy to give a good
indication of the intrinsic vortex--vacancy interaction potential.

What this means in practice is that, with the vortex fixed at the origin
(center of circle of radius $R$), all the lattice sites are allowed to
shift as the vacancy takes on a range of positions.
This implies that lattice sites periodically pass by the vortex
as the vacancy is moved to a sequence of positions.
To avoid an undefined in-plane spin direction, a lattice site 
should not be allowed to fall squarely on the vortex position;
as a result, some choices of vacancy position are prohibited.

For this numerical algorithm, it is simple to specify the vacancy position,
which influences the positions of the rest of the lattice sites.
On the other hand, an exact specification of a desired vortex 
position is not possible, because the presence of an off-center
vacancy near a vortex skews the vortex spin field, leading to
an ambiguity in the position of the ``vortex center.''
Indeed, without a sufficient constraint, if an initial nearly in-plane
vortex is placed near a vacancy, and then the relaxation procedure already
described (Sec.\ \ref{relaxed}) is applied, it is possible that the 
vortex simply shifts onto the vacant site.
Therefore, a method is needed to at least partially enforce some
desired vortex position.

The following simple but somewhat arbitrary scheme was applied to 
enforce a desired vortex core position.
Based on the arrangement of the lattice sites near the vortex,
which is not necessarily symmetrical, a few spins nearest to the 
vortex are considered as ``core'' sites.
The in-plane angles of these core sites are not changed 
by the energy relaxation scheme using Eq.\ (\ref{Field}).
Their $\Phi_{\bf n}$ are held fixed at the original angles given according 
to the in-plane profile, Eq.\ (\ref{Phi_in}), however, their out-of-plane
spin components {\em are} allowed to change.  
The number of core sites is taken to be $n_c=6, 4$ or $3$ for 
hexagonal, square or triangular lattices, respectively.
While it is somewhat artificial and arbitrary to constrain some
spins, it was found to be necessary to maintain the desired vortex
position, especially at vortex to vacancy distances less than a few
lattice constants. 
This restriction of some spin movement may affect the final energy,
therefore one can only consider this as an estimate of the
vortex--vacancy potential.
However, at a position of high symmetry, such as the vortex on top
of the vacancy, the constraint has no effect.
%

Obviously this is a rich problem with a wide choice of parameters.
One expects the potential to be influenced strongly by the choice
of anisotropy constants.
Here only systems with exchange anisotropy are studied; 
one expects the results for single-ion anisotropy to be comparable.
Furthermore, there could be notably different behaviors depending on
where the chosen anisotropy constant $\lambda$ lies relative to
the vortex-on-plaquette critical anisotropy $\lambda_c$ and the
vortex-on-vacancy critical value $\lambda_{cv}$. 
A particularly interesting and complex case is when $\lambda$ lies between 
$\lambda_c$ and $\lambda_{cv}$; the vortex will take an out-of-plane
form far from the vacancy, and transform into a planar vortex as
it is set closer to the vacancy.
For $\lambda$ smaller than both $\lambda_c$ and $\lambda_{cv}$, the vortex
always remains planar.
Conversely, for $\lambda$ larger than both $\lambda_c$ and $\lambda_{cv}$, 
the vortex always remains out-of-plane, although the magnitude of its
out-of-plane moment will change as it moves onto the vacancy position,
as already indicated in Table \ref{Reduction}.
%

\subsubsection{Vortex Remaining in a Planar State}
\begin{figure}
\psfig{figure=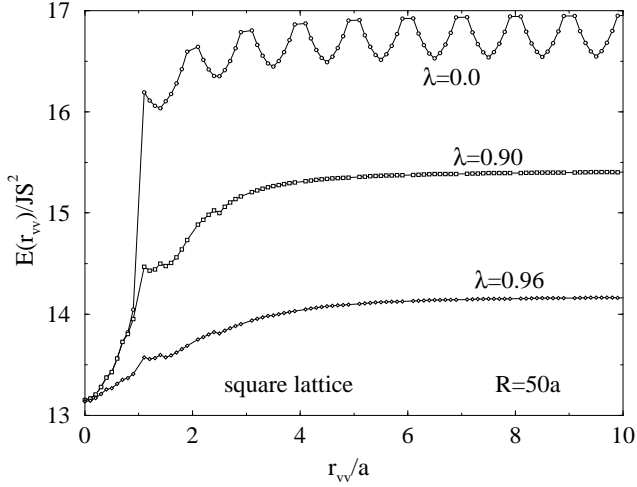,angle=-90.0,width=\pssize pt}
\caption{
\label{sqr50Pot}
Vortex--vacancy total energy as a function of their separation $r_{\rm vv}$,
calculated on a square lattice system of radius $R= 50a$ for indicated
exchange anisotropies ($\lambda=0$, in-plane; $\lambda=0.90$, transition
from in-plane to out-of-plane with increasing $r_{\rm vv}$; $\lambda=0.96$,
out-of-plane vortices).  With the vortex at the origin $(0,0)$, the vacancy
was placed at a sequence of positions in the (10) direction.
}
\end{figure}

For sufficiently strong anisotropy ($\lambda<\lambda_{c})$, the vortex 
always remains planar at any vortex--vacancy separation.
The parameter choice $\lambda=d=0$ was used to obtain this. 
A typical result for the energy is shown in the uppermost curve 
of Fig.\ \ref{sqr50Pot}, for a square lattice system, with radius $R= 50a$.
Curves with essentially the same shapes were obtained also on a system
with $R=30a$, except for an overall shift to lower energies due to
the usual logarithmic dependence on system size.
One expects that except for fine details (such as the width or depth of
the vacancy--vortex potential), the gross features of the potential
interaction should not depend on the choice of the lattice (the usual 
assumption of continuum theory).

The curves in Fig.\ \ref{sqr50Pot} correspond to vacancy positions
along paths in the (10) direction of the lattice, avoiding those points 
which would cause the vortex to fall exactly on a lattice site.
A fairly deep attractive minimum ($\Delta E \approx 3.3 JS^2$) is found for 
the vacancy centered on the vortex, with obvious periodic wings due to the 
lattice.
Similar results are obtained for vortex--vacancy displacements along other
directions, but with a slightly different width of the potential minimum
(the vortex--vacancy potential is not isotropic).
Clearly this confirms the energy deviation found in Sec.\ \ref{DeltaE},
and furthermore, gives an indication of the radial separation over which the
vacancy has a strong influence on a vortex (1--2 lattice constants).
The periodic variations for large vortex--vacancy separation demonstrate
the relatively weaker local potential effects that tend to center a 
vortex within an individual plaquette.
If one did not constrain the small set of core spins to enforce a
desired vortex position off-center in a plaquette, then the relaxation
would simply produce a vortex profile centered in the nearest convenient 
plaquette, and these periodic variations would be greatly reduced.
Similar results hold for the hexagonal and triangular lattices, but with
potential depths as found in Sec.\ \ref{DeltaE}, and periodic 
variations in the wings that are distinctive to the lattice.
When $\lambda<\lambda_{c}$, the total out-of-plane magnetization remains at 
zero for the whole range of vortex--vacancy separations.

\subsubsection{Vortex Remaining in an Out-of-Plane State}
It was found above that for out-of-plane vortices, the pinning
energy was much weaker than for in-plane vortices. 
Thus it makes sense also to compare the radial dependence of the 
vortex-vacancy potential interaction for out-of-plane vortices to 
that for in-plane vortices.

\begin{figure}
\psfig{figure=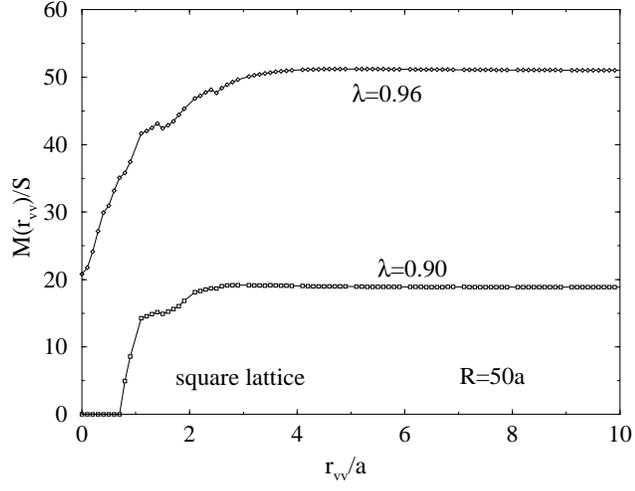,angle=-90.0,width=\pssize pt}
\caption{
\label{sqr50M}
Vortex--vacancy total out-of-plane magnetization as a function of separation 
$r_{\rm vv}$, for square lattice systems of radius $R=50a$ as in 
Fig.\ \protect\ref{sqr50Pot}.  ($M=0$ for all $r_{\rm vv}$ at $\lambda=0$.)
}
\end{figure}

As an example of this possibility, consider exchange anisotropy constant
$\lambda=0.96$ for the square lattice, larger than the value 
$\lambda_{cv} \approx 0.9545$, which results in stable out-of-plane 
vortices regardless of their position. 
The potential curve for separations along the (10) direction of
the lattice is shown in the lowermost curve of Fig.\ \ref{sqr50Pot}.
Only slightly different potential curves result for separations along
other directions. 
The potential curve is very smooth compared to that for the in-plane
vortices ($\lambda=0$), which must be due to the relatively low
energy out-of-plane tilting adjustments of the spins.
The corresponding change in the overall system out-of-plane magnetization 
is shown in Fig.\ \ref{sqr50M}.
As the vortex is moved closer to the vacancy, its out-of-plane magnetization
diminishes by $\Delta M\approx 30 S$ at the same time that its energy is 
mildly reduced by $\Delta E \approx 1.0 JS^2$, with an overall attraction. 
Of course, the depth of the attraction will depend on the particular 
choice of anisotropy strength.

If this type of situation holds in a magnetic material,
(large change in $M$ associated with small energy difference)
there could be large out-of-plane magnetization fluctuations
expected as vortices are alternately pinned and freed by vacancies,
due to thermal fluctuations.
The degree of this effect will be greatest for an anisotropy parameter
$\lambda$ approaching the isotropic limit, $\lambda\approx 1$,  
corresponding to the largest magnetization {\em change} together with the
smallest energy difference between the free and pinned vortex states.
The two different states, however, could be difficult to distinguish in
a real medium, because both have nonzero (and large) $M$.

\subsubsection{Vortex Transforming From Out-of-Plane to In-Plane}
Alternatively, consider exchange anisotropy $\lambda=0.90$ for the 
square lattice, which is above $\lambda_c \approx 0.7034$ and 
below $\lambda_{cv} \approx 0.9545$.
The vortex--vacancy potential is shown as the middle curve of 
Fig.\ \ref{sqr50Pot}.
At this anisotropy strength, the magnetic moment for the separated
out-of-plane vortex is about $M\approx 19 S$.
As the vortex is pulled closer to the vacancy, it minimizes its energy
by reducing the out-of-plane magnetization to zero, transforming 
dramatically from out-of-plane form to in-plane form (Fig.\ \ref{sqr50M}).
In the process, one sees an intermediate overall energy reduction of about 
$2.3 JS^2$.
If such a transition between the pinned planar and free out-of-plane 
states could be controlled externally by some applied field, such a 
pair of states could have practical applications for data storage,
especially because of the transition to an easily distinguished
zero magnetization state.

\section{Conclusions}
A vacancy has been found to have substantial influence on the structure
and stability properties of a vortex in an easy-plane magnet.
The quasi-static analysis presented here is valid for FM and AFM models 
on square and hexagonal lattices, and only for the FM model on a
triangular lattice.
Consistent with the stability results found by Zaspel et al,\cite{Zaspel96}
when a vortex is energetically relaxed centered on a vacancy, relatively 
weaker easy-plane anisotropy will stabilize it in the planar form, compared 
to that necessary when the vortex is centered in a plaquette far from
any vacancy.
This result holds as well on hexagonal and triangular lattices; 
the critical anisotropy strengths $\delta_{cv}$ and $d_{cv}$
are always reduced for vortex-on-vacancy compared to those for
vortex-in-plaquette, $\delta_c$ and $d_c$.
All the critical anisotropy strengths are smallest on the hexagonal lattice 
and largest on the triangular lattice, which suggests that the coordination
number and lattice density control the effect.

Using a simple scheme to enforce a desired vortex position,
energetically relaxed vortex--vacancy configurations were produced.
The potential interaction between the two has been estimated for the 
different lattices and anisotropy regimes.
In all cases studied, the vortex--vacancy interaction is attractive.
This is important because it establishes the vortex-on-vacancy
states as legitimate local minimum energy configurations.
The depth of the potential holding a vortex on a vacancy (pinning energy)
has been estimated. 
It tends to be strongest on the triangular lattice and weakest on the 
hexagonal lattice for in-plane vortices.
For out-of-plane vortices, the typical depth of the vortex pinning potential
tends to be much weaker than for in-plane vortices.
%

An interesting effect is expected for a system with anisotropy
strength intermediate between the vortex-in-plaquette and vortex-on-vacancy
critical anisotropies (for example, $\lambda_{c} < \lambda < \lambda_{cv}$).
In the transformation from out-of-plane to in-plane type as a
vortex moves from free to pinned on a vacancy, the out-of-plane
magnetization is completely eliminated.
This could lead to strong magnetization fluctuations,
and also be considered as an additional signature of
vortices and a distinct signature of the pinning process.
Additionally, the magnetic fluctuations associated with pinning and 
de-pinning of {\em out-of-plane} vortices may be even stronger: 
relatively large out-of-plane magnetization changes will occur at very 
low energy difference between the pinned and free vortex states.
The largest magnetization changes in the pinning process occur at 
the vortex-on-vacancy critical anisotropy, $\lambda=\lambda_{cv}$. 

Alternatively, one could consider application of a weak out-of-plane
magnetic field, which would bias the out-of-plane polarizations of 
generated vortices along that preferred direction 
(generating the ``light'' cone-state vortices\cite{IvSheka95,IvWysin02}).
An initial low-temperature state of planar vortices pinned on vacancies
could produce de-pinned out-of-plane vortices preferentially with the same
polarization, leading to a large macroscopic magnetic moment.
This could be a intriguing effect if the de-pinning and pinning of vortices 
on vacancies could be controlled at will, leading to interesting device 
possibilities.

Finally, one can speculate on how an attractive potential between vacancies and
vortices should affect the statics and dynamics of the BKT topological transition.
The MC results by Leonel et al\cite{Leonel+03} indicated that 
a small percentage of spin vacancies lowers the transition temperature, 
which they interpreted to result from a {\em repulsive} vacancy--vortex potential
(under an assumption of a certain vortex deformation caused by a vacancy).
The explanation requires that the repulsion produces a greater free-vortex
density at a lower temperature, thereby leading to more disorder and a
lower transition temperature. 
Actually, we found here that the vacancies are likely to {\em attract and pin} any 
vortices/antivortices that are forming, with fairly strong pinning energies 
for the planar model, and associated lower vortex formation energies.
An improved continuum model for vacancy--vortex interaction, without the
assumption of a global vacancy-induced vortex deformation, has 
verified this attractive potential.\cite{Afranio03}
At temperatures well below the energy scale of $JS^2$, nearly all formed
vortices should be pinned.  
Even so, the pinned vortices will lead to long range disorder and the BKT 
transition, but at a lower temperature, especially because their formation energy 
is {\em much lower} than for unpinned vortices.  
Further detailed MC studies could be interesting if they could demonstrate 
the respective roles of the free versus pinned vortices, by analyzing their
number densities.
This question may also be answered by studying the dynamic correlations:
the dynamic signatures of free vortices should be distinct from those
due to pinned vortices.
%


\ifnum\pageformat=1
  \newpage
\fi

\end{document}